\documentstyle[12pt,aasms4]{article} 

\received{10/24/97}
\lefthead{}
\righthead{}
\slugcomment{ApJ subm 10/24/97}

\def\hl{\hline}
\def\mathnew{\mathsurround=0pt}
\def\simov#1#2{\lower .5pt\vbox{\baselineskip0pt \lineskip-.5pt
        \ialign{$\mathnew#1\hfil##\hfil$\crcr#2\crcr\sim\crcr}}}
\def\simg{\mathrel{\mathpalette\simov >}}

\def\be{\begin{equation}}
\def\ee{\end{equation}}
\def\bc{\begin{center}}
\def\ec{\end{center}}
\def\L{\Large}
\begin{document}

\title {A 3$^{rd}$ CLASS OF GAMMA RAY BURSTS?}

\author{ I. Horv\'ath }
\affil{Department of Astronomy \& Astrophysics, Pennsylvania State University,
525 Davey Laboratory, University Park, PA 16802, and Dept. of Earth Science,
Pusan National University, Pusan 609-735, 
Korea, e-mail: hoi@astrophys.es.pusan.ac.kr }
%
%\centerline{To appear in {\it The Astrophysical Journal}}
%\centerline{\it Received 1997 Oct 24; accepted ___}
%
\begin{abstract}
Two classes of Gamma Ray Bursts have been identified so far, characterized
by $T_{90}$ durations shorter and longer than approximately 2 seconds. We show
here that the BATSE 3B data allow a good fit with three Gaussian distributions
in $\log T_{90}$. The $\chi  ^2$ statistic indicates a 40 \% probability for
two Gaussians,  whereas the three-Gaussian fit probability is 98 \%.  Using 
another statistical method, it is argued that the probability that the third 
class is a random fluctuation is less than 0.02 \%.
\end{abstract}
\section{Introduction}
In the BATSE 3B catalog (\cite{M6}) there are 1122 Gamma-Ray Bursts (GRBs), 
of which 834 have duration information. 
\cite{K1} have identified two types of GRB based on durations, 
for which the value of $T_{90}$ (the time during which 90\% of the fluence 
is accumulated) is respectively smaller or larger than 2 s. This bimodal
distribution has been further quantified in another paper (\cite{K}),
 where a two-Gaussian fit is made. In order to make further progress
in quantifying this classification, one of the issues which needs to be 
addressed is an evaluation of the probabilities associated with a bimodal,
or in general multimodal distribution. In this paper we take a first
attempt at this, evaluating the probability that the two populations are
independent, and we consider in addition whether a third group of bursts
can be identified.
\section{ GAUSSIAN FITS IN $\log T_{90}$} 
For this investigation we have used a smaller set of 797 burst durations in 
the 3B catalog, because these have peak flux information as well, and we 
use the T$_{90}$ measures provided in this data set.
For a one-Gaussian fit the $\chi  ^2$ probability is less than 0.1 \%, so
the one-Gaussian null hypothesis can be rejected. A log-normal two-Gaussian 
fit using 52 time bins with bin size 0.1 in the log (fig. 1.) has a 40 \% 
probability. This indicates that this null hypothesis cannot be rejected 
(\cite{P}). However, in the middle we have three consecutive bins, 
which have extremely large deviations.

The interesting three bins in the middle have  3.2 $\sigma $, 4 $\sigma $ and 
2.5 $\sigma$ deviations (fig. 2.). The probability to get such deviations in 
52 bins is 6.3 \% , 0.3 \% and 58 \% . Because we obtained these in three 
consecutive bins the probability that these deviations are random is 
$\sim $ 10$^{-4}$.

Such a deviation is interesting if it is in one/two/etc.
bins. This raises the probability approximately by a factor five.
These three bins contain 64 GRBs. The two-Gaussian fit would account for 19,
leaving 45 extra. The difference between these two numbers is more than 
5 $\sigma $, which corresponds to a probability of less than 
$6\times 10^{-7}$.
We can ask what is the probability of an excess in n (n small) consecutive 
bins located somewhere else. The number of trials is of order 
the number of possible locations of excess {\it times} the number of 
possible widths of excess, which is roughly 
50 $ \times $ 6 = 300.
Therefore if the GRB duration distributions are log-normal then the
probability that there is a deviation in these three bins from a 
two-Gaussian fit is more than 99.98 \%. 

Because the duration distribution of the previously known two classes of GRB 
are nearly log-normal, we made a fit including a third Gaussian.
The three-Gaussian fit (with nine parameters) has a $\chi  ^2$ = 24, which 
implies a 98 \% probability (fig. 3.). Table 1. contains the parameters of
this fit. Looking at the goodness of the three-Gaussian fit, one might 
conclude that there is an ``intermediate" type of GRB.
This is an agreement with the  \cite{xx} result, who used a
multivariate analysis and find that the probability of existence of two 
clusters rather than three is less than 10$^{-4}$.

% First change
{
Both the two- and three-Gaussian fits have acceptable $\chi  ^2$ values, 
therefore neither can be rejected as an inappropriate fit. If we view the 
problem as one of model comparison, then, if the third Gaussian does not 
exist, the change in $\chi  ^2$ between the two fits from adding the 
three-parameter third Gaussian should be distributed as $\chi  ^2$ of three 
degrees of freedom (\cite{B}). The change in $\chi  ^2$ of 46.8-24.0=22.8 
implies a chance probability of less than $10^{-4}$ indicating that a three 
Gaussian fit is a highly significant improvement over a two Gaussian fit.
}

If we had considered six bins in the ``intermediate" region, the two-Gaussian
fit would show an excess in these bins of 61 GRBs. We may estimate that
the ``intermediate" group contains more than 50 but less than 70 GRBs, which 
would represent $\sim 8$ \% of the GRB population.
\section{ SYSTEMATICS}
The BATSE on-board software tests for the existence of bursts by comparing 
the count rates to the threshold levels for  three separate time intervals: 
64 ms, 256 ms, and 1024 ms. The efficiency changes in the region of the middle 
area because the 1024 ms trigger is becoming less sensitive as burst 
durations fall below about one second. This means that at the ``intermediate"
timescale a large systematic deviation is possible. 
To reduce the effects of trigger systematics in this region we 
truncated the dataset to include only GRBs that would have triggered BATSE 
on the 64 ms timescale.

Using the Current BATSE catalog CmaxCmin table (\cite{M}) we choose the GRBs,
which numbers larger than  one in the second column (64 ms scale
maximum counts divided by the threshold count rate).
Because this process reduced the bursts numbers very much
we used the 4B catalog (\cite{Pa}) data.
In the 4B catalog there are 1234 GRBs which have duration information; 
unfortunately just 605 burst satisfied the above condition.
In the interesting three bins we have 35 GRBs.
The two-Gaussian fit  would account for 13,
leaving 22 extra. Therefore 63 \% of the population are still
a ``deviation" (note in Sect. 2. it was 70\%).

This  is still a  4-5 $\sigma $ deviation, depending on whether one uses
the expected number or the observed one. 
Therefore after neglecting  some systematics the probability 
that there is a deviation in these three bins from a two-Gaussian
fit is about 99.8 \% or 99.98 \%.
\section{ LOG N - LOG P  }
In Fig. 4 we show the $\log N - \log P$ distribution of the two previous
classes of GRB (see also \cite{K} and \cite{HMM}), 
and compare this with the $\log N- \log P$ in Fig. 5 for the ``intermediate" 
class of bursts, consisting of 74 events (logT$_{90}$=0.4-0.8).
The logN-logP distributions of the T$_{90}<$2.5 sec and the T$_{90}>$6.3 sec
are on the fig. 4. 
The former has a Euclidean part, and the left portion goes over into a -1.05 
slope. The second group has a -1.06 slope.
This corresponds to the results of \cite{K}.

The logN-logP distribution of the ``intermediate" type  GRBs
% Second change
{
(fig. 5.) may be different, but this is questionable.
}
\section{ HARDNESS RATIOS }
The average hardness ratios of the short and long bursts are 4.5 and 2.5. The 
``intermediate" type has 2.2 average hardness ratio. If this group were
a random mix of some {\it shorter} long-type bursts and some {\it longer} 
short-type of bursts, the expected value should have been between the two 
average hardness ratios. The actual value differs significantly from this, 
being close (and even somewhat smaller) than the hardness ratio of the long 
bursts. However, from a purely hardness-ratio point of view, it does not
seem possible to distinguish the ``intermediate" bursts from the long ones,
thus the hardness information does not support, but also does not contradict,
the ``intermediate" burst class hypothesis.

\section{ CONCLUSION} 
It is possible that the three log-normal fit is accidental, and that there 
are only two types of GRB. However, if the T$_{90}$ distribution of 
these two types of GRBs is log-normal, then the probability that the third 
group of GRBs is an accidental fluctuation is less than 0.02 \%.
The logN-logP distribution and hardness ratio are also suggestive of the fact
that the intermediate duration (T$_{90}$=2.5-7 sec) bursts represent a third
class of GRBs.

\acknowledgments

This research was supported in part through NASA5-2857, OTKA T14304,
Sz\'echenyi Fellowship and KOSEF. Useful discussions with E. Feigelson,
E. Fenimore, A. M\'esz\'aros, P. M\'esz\'aros, and J. Nousek are appreciated.
The author also thanks an anonymous referee for useful comments that
improved the paper.

\clearpage

\bigskip
\bigskip

Figure caption 1.

Distribution of $\log T_{90}$ for 797 bursts from the 3B catalog. The solid 
line represents a fit of two log-normal Gaussians using 6 parameters and 52 
bins. The best fit $\chi^2$=46.8, which implies a 40 \% probability.
\bigskip

Figure caption 2.

Difference between the two log-normal Gaussian fit and the $\log T_{90}$
distribution. The solid lines represent one sigma errors of the T$_{90}$ bins.
Dotted lines and dashed lines mean two and three sigmas. Except for the three 
middle bins, only one bin has $\simg 2$ sigma deviation, and seven bins have
$\simg 1$ sigma.
\bigskip

Figure caption 3.

$\log T_{90}$ distribution. The solid line represents a fit of three 
log-normals. The $\chi^2$ value implies a probability of 98 \% .
\bigskip

Figure caption 4.

The logN-logP distributions of the two type of GRBs. The left curve is for
T$_{90}<$2.5 sec (223 GRBs), while the right curve is for T$_{90}>$6.3 sec 
(508 GRBs).
\bigskip

Figure caption 5.

logN-logP distribution of the ``intermediate" type (2.5$<$T$_{90}<$6.3) GRBs. 
The slope of -0.65 is substantially different from that of the two other 
types of GRBs. This is compatible with the "intermediate" GRB being a 
separate class of objects.

\clearpage

\bc {\L Table 1.} 

 $$
\begin{array}{rrrrccr}\hl\hl
&&&&&\cr
 & mid (\lg s) &\sigma (\lg s) &  \ memb & \% & T_{90} \ \ 
\cr\hl\hl
first \ \ & - \ 0.35 & 0.50 & 236 \ \ &\ 30  &   < 5 \ s \   \cr
 second   & 1.52 & \ \ 0.37 & 497\  \ &\ 62 & > 4 \ s \    \cr
third \ \  & 0.64 & 0.14 & \  61 \ \ &\ \ \ 8 & 2.3-8   
     \cr\hl\hl
\end{array}
$$  \ec

\vskip 2.2 truecm

The parameters of the three-Gaussian fit of GRBs. 

Average duration  ($mid$),
width of the Gaussian ($\sigma $) (both in lg $sec.$).
Members of the group ($memb$) and population of the group (\%).
Border of the group in $T_{90}$ (unfortunately they overlap).

\end{document}